\documentclass[sigconf]{acmart}
\renewcommand\footnotetextcopyrightpermission[1]{} 

\AtBeginDocument{%
  \providecommand\BibTeX{{%
    Bib\TeX}}}

\setcopyright{acmlicensed}
\copyrightyear{2027}
\acmYear{2027}
\acmDOI{}
\acmConference[KDD '27]{ACM SIGKDD Conference on Knowledge Discovery and Data Mining}{August 2027}{San Jose, California}
\acmISBN{}

\usepackage{textcomp}
\usepackage{verbatim}
\usepackage{amsfonts}
\usepackage{amsmath}
\usepackage{array}
\usepackage{subfigure}
\usepackage{caption}
\usepackage{balance}
\usepackage{multicol}
\usepackage{multirow}
\usepackage{color}
\usepackage{epsfig}
\usepackage{natbib}
\usepackage{xspace}
\usepackage{booktabs}
\usepackage{bm}
\usepackage{enumitem}
\usepackage{diagbox}
\usepackage{url}
\usepackage{geometry}
\usepackage{changepage}
\usepackage{textcomp}
\usepackage{graphicx}
\usepackage{color}
\usepackage{bm}
  
\usepackage{amssymb}
\usepackage{algorithm}
\usepackage{algorithmic}
\usepackage{threeparttable}
\usepackage{mathrsfs}
\usepackage{soul}
\usepackage{tcolorbox}
\usepackage{listings}
\tcbuselibrary{breakable, skins}

\newcommand{\vpara}[1]{\vspace{0.05in}\noindent\textbf{#1 }}
\newcommand{\vvpara}[1]{\vspace{0.05in}\noindent\textit{#1 }}
\newcommand{\model}{DeepScrub\xspace}

\usepackage{listings}

\begin{document}

\title{Traceable LLM Reasoning for Fake-Order Fraud Detection}

\author{Siqi You}
\affiliation{%
  \institution{ByteDance}
  \city{Hangzhou}
  \country{China}}
\email{yousiqi.147@bytedance.com}

\author{Bingsong Xu}
\affiliation{%
  \institution{ByteDance}
  \city{Beijing}
  \country{China}}
\email{xubingsong@bytedance.com}

\author{Zhixian Zheng}
\affiliation{%
  \institution{ByteDance}
  \city{Hangzhou}
  \country{China}}
\email{zhengzhixian@bytedance.com}

\author{Xinjian Peng}
\affiliation{%
  \institution{ByteDance}
  \city{Hangzhou}
  \country{China}}
\email{pengxinjian@bytedance.com}

\author{Yang Xie}
\affiliation{%
  \institution{ByteDance}
  \city{Hangzhou}
  \country{China}}
\email{xieyang.bd@bytedance.com}

\author{Ying Wang}
\affiliation{%
  \institution{ByteDance}
  \city{Hangzhou}
  \country{China}}
\email{wangying.826@bytedance.com}

\author{Jiarong Xu}
\affiliation{%
  \institution{Fudan University}
  \city{Shanghai}
  \country{China}}
\email{jiarongxu@fudan.edu.cn}

\renewcommand{\shortauthors}{You et al.}








\begin{abstract}
Detecting fake-order fraud at scale remains a critical challenge for large online-to-offline (O2O) service platforms, as existing approaches often rely on expert-designed features, produce black-box decisions, and provide limited interpretability. To address these limitations, we propose \model\footnote{The name DeepScrub reflects the goal of deep-cleaning fake-order transactions from e-commerce platforms.}, a reinforcement learning framework built upon large language models (LLMs) for fake-order fraud detection with traceable reasoning. \model introduces three innovations. First, a semantic unification module converts heterogeneous risk signals into textual descriptions that LLMs can understand. Second, continued pre-training on risk-control corpora injects domain knowledge, and task rewards jointly evaluate prediction correctness and reasoning quality. Third, the SUggest-REflect (SURE) mechanism incorporates expert feedback and model self-checking to iteratively refine reasoning paths. On a real-world fake-order fraud detection dataset, \model achieves a macro-F1 score of 85.3\%, outperforming the best baseline by 2.7 percentage points. Our task-optimized 8B model further surpasses a 32B model, showing that domain adaptation can matter more than model scale in this setting. In a four-week live pilot, \model achieved 91.8\% precision and 88.5\% recall, improving over first-stage human reviewers by 16.6 and 38.8 percentage points. It reduced first-stage manual review workload by 94\% and saved nearly one million RMB annually. These results show that \model improves fraud review accuracy, reduces first-stage review workload, and provides traceable evidence for production risk-review workflows.
\end{abstract}

\maketitle

\section{Introduction}
\label{sec:intro}

Fake-order fraud, also known as brushing, refers to behaviors that fabricate or manipulate transaction records, clicks, or user reviews to create an illusion of inflated commercial scale, popularity, or user engagement. 
Such practices not only undermine the reliability of sales volumes and ratings, but also distort fair competition among merchants, posing substantial risks to the socio-economic system. 
According to statistics from the Ministry of Public Security of China in 2025, brushing-related fraud accounts for about 37\% of all telecommunication fraud cases. 

Controlling fake-order fraud on large O2O platforms is particularly challenging because suspicious orders are interleaved with legitimate business bursts, promotional traffic, and diverse merchant operating strategies. With hundreds of millions of daily transactions and complex multi-party interactions, purely manual review is costly and difficult to scale. At the same time, purely model-driven decisions are often hard to justify in enforcement and appeal workflows, where platforms must explain why a merchant or transaction is judged to be suspicious.

Existing fake-order fraud detection methods mainly fall into two categories. The first category relies on handcrafted features and expert rules derived from operational signals and historical behavior patterns. Although effective in some high-precision scenarios, these methods require heavy expert involvement, incur high maintenance costs, and adapt poorly to evolving fraud strategies. The second category uses deep learning models to improve detection accuracy, but their black-box predictions make it difficult to provide defensible evidence in real enforcement workflows.

 This limitation is especially serious in local-service O2O scenarios, where online orders are tightly coupled with offline fulfillment. In such settings, enforcement decisions are not merely classification outputs; they may directly affect merchant penalties, appeal resolution, and downstream operational actions. For example, a legitimate multi-branch merchant may exhibit frequent transactions, recurring users, or unusual geographic patterns that superficially resemble brushing behavior. Without business-grounded explanations, platforms risk penalizing legitimate merchants and weakening trust in the review process. Therefore, an effective fake-order fraud detection model should not only identify suspicious transactions, but also provide traceable evidence that supports review, enforcement, and appeals.

This requirement makes large language models (LLMs) appealing for fake-order fraud detection. By reasoning over heterogeneous evidence in natural language, LLMs can potentially connect risk signals to human-readable explanations and produce auditable decision traces. However, directly applying LLMs to this setting is still insufficient for deployment. In practice, three challenges remain.

\textbf{First, key evidence for fake-order fraud is distributed across heterogeneous data sources, including association graphs, behavioral sequences, and structured transaction records, whereas LLMs primarily consume text.} A deployable system therefore needs a practical way to unify heterogeneous evidence into a representation that supports both reasoning and auditing without introducing excessive system complexity. \textbf{Second, fake-order fraud detection requires optimizing not only prediction correctness but also reasoning quality}, because correct labels with weak or unsupported explanations are still inadequate for enforcement workflows. \textbf{Third, high-stakes fraud review cannot rely solely on a single-pass generation process.} When initial reasoning overlooks key merchant-side or transaction-side evidence, the system should be able to revisit and refine its judgment in a structured manner.

To address these challenges, we develop \model, an LLM-based reinforcement learning framework for fake-order fraud detection. \model first converts heterogeneous risk signals into unified textual evidence through a semantic unification module, enabling the model to reason over graphs, sequences, and structured records in a common format. It then performs continued pre-training on risk-control corpora to inject domain knowledge, followed by task-specific reward design to optimize both prediction correctness and reasoning quality. Finally, we introduce the SUggest-REflect (SURE) mechanism that combines expert feedback and self-verification to refine intermediate reasoning and improve the reliability of final decisions. The resulting model improves detection performance and provides traceable evidence for production risk-review workflows.

We evaluate \model on a real-world fake-order fraud detection dataset collected from a large O2O platform. The proposed framework achieves the best performance among the evaluated methods in offline experiments. More importantly, in a four-week live pilot, \model achieved 91.8\% precision and 88.5\% recall, outperforming human reviewers by 16.6 and 38.8 percentage points, respectively, while reducing manual review workload by 94\% and saving nearly one million RMB annually. These results demonstrate that LLM-based reasoning can be translated into measurable operational value in real-world risk-review settings.

In summary, our key contributions are as follows:

\begin{itemize}[leftmargin=*]
    \item We present \model, an LLM-based reinforcement learning framework for fake-order fraud detection in production risk-review workflows on large O2O platforms.
    \item We combine semantic unification, continued domain adaptation, task-specific reward modeling, and the SUggest-REflect (SURE) mechanism to improve prediction accuracy and reasoning quality.
    \item We evaluate \model through offline experiments and a four-week live pilot, showing improved review accuracy, a 94\% reduction in first-stage manual review workload, and nearly one million RMB in annual cost savings.
\end{itemize}

\begin{figure*}[h]
  \centering
  \includegraphics[width=0.8\linewidth]{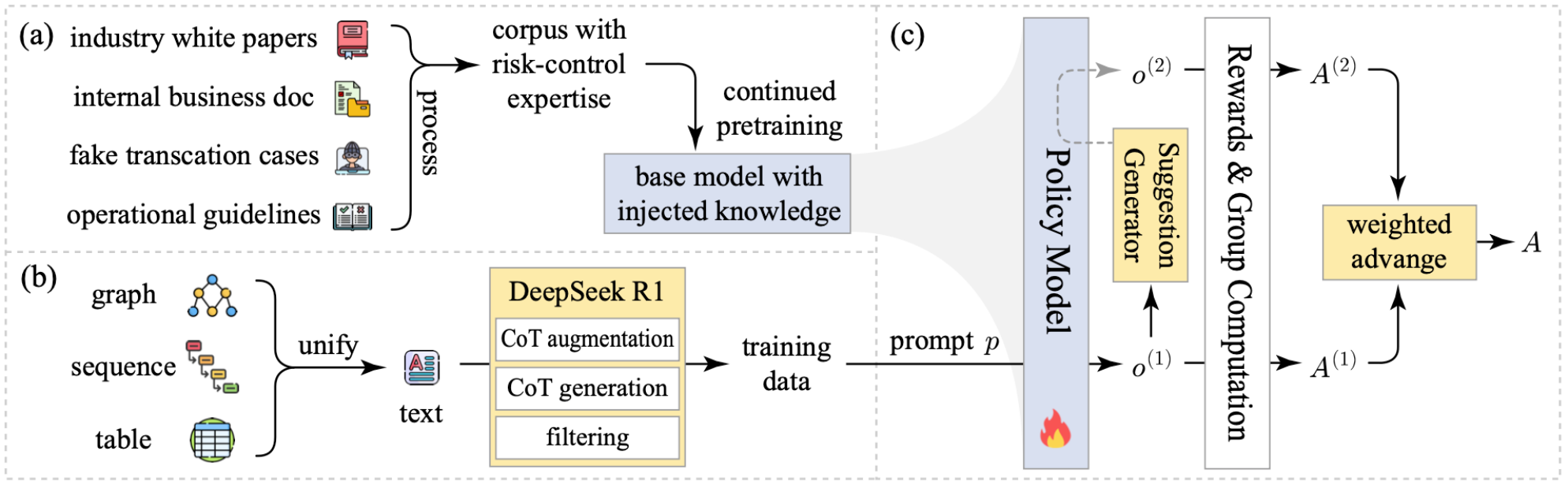}
  \vspace{-3mm}
  \caption{
  Overview of \model. The system first adapts a base LLM with risk-control corpora, then converts heterogeneous order evidence into text, and finally applies SURE training to optimize both fraud labels and reasoning traces.
  }
  \label{fig:DeepScrub}
\end{figure*}

\section{Related Work}
\vpara{Fraud Detection}
Fraud detection has been widely studied in e-commerce, finance, and online service platforms. Early and widely deployed approaches rely on expert rules and manually designed risk indicators, which remain effective for high-precision screening and operational recall~\cite{thomas2017credit,dheepa2012behavior,del2017enhancing}. Machine learning methods further improve detection by learning behavioral and transactional patterns from historical data~\cite{app16020876,ali2022financial}. Since many fraud activities are organized through linked accounts, merchants, or coordinated communities, graph-based fraud detection has also become an important direction~\cite{pourhabibi2020fraud,cheng2025graph,pan2025label}. More recently, LLMs have been explored for anomaly and fraud detection over tabular or transaction-level data~\cite{tsai2025anollm,wang2026causaltad,cheng2025llmtad}. These methods are useful for risk scoring, filtering, and anomaly detection, but they usually provide limited natural-language reasoning that can be inspected during enforcement, escalation, or appeal handling. This limitation motivates our focus on fake-order fraud review, where the system must produce both a decision and a reviewable reasoning trace.

\vpara{Reinforcement Learning in LLMs}
With the development of large models in various practical applications~\cite{zhang2026moon,fu2025moon,nie2026moon2,wu2026moon3}, reinforcement learning (RL) has also become a common method for improving and adjusting the behavior of large language models (LLM).
Reinforcement learning from human feedback (RLHF) trains reward models from human preferences and then optimizes language models with reinforcement learning algorithms such as REINFORCE and proximal policy optimization (PPO)~\cite{christiano2017deep,williams1992simple,schulman2017proximal,ouyang2022training}. Later preference-optimization methods, including KTO and DPO, reduce the complexity of reward modeling while improving alignment efficiency~\cite{ethayarajh2024kto,rafailov2023direct}. group relative policy optimization (GRPO) further improves training efficiency by removing the need for a separately trained critic model~\cite{guo2025deepseek}. These methods mainly optimize general preference alignment or final-answer correctness. Fake-order fraud review requires a more specific objective. The model must generate a correct label and a reasoning trace grounded in transaction evidence. \model addresses this need by combining domain adaptation, task-specific rewards, and SURE training for a deployed review workflow.
\section{Problem Formulation}
\vpara{Fake-order Fraud Detection.}
Given a de-identified textual representation of an order, the task is to determine whether the order is involved in fake-order fraud and identify its underlying motivation. The input integrates heterogeneous evidence, including relational networks, temporal behavior sequences, and structured transaction records. The model outputs a reasoning trace $e$ and a label $y \in \mathcal{Y}$. The label space is $\mathcal{Y}=\{0,1,2\}$, corresponding to normal transaction, merchant inflating product sales, and influencer inflating their level.
\section{Method of \model}

\begin{table}[b]
\caption{Results under different pretraining data ratios.}
\vspace{-3mm}
\label{knowledge}
\centering
\begin{tabular}{lcc}
	\toprule
	Data & C-Eval & C-MMLU \\
	\midrule
	Qwen3-8B & 73.87 & 76.93 \\
    \midrule
	Qwen3-8B (only R) & 69.04 & 73.44 \\
	Qwen3-8B (R:G=1:5) & 70.59 & 74.23 \\
	Qwen3-8B (R:G=1:10) & 72.86 & 75.11 \\
	Qwen3-8B (R:G=1:15) & \textbf{73.97} & \textbf{75.80} \\
	\bottomrule
\end{tabular}
\end{table}

\subsection{Overview}
\label{sub:overview}

In this section, we describe the technical details of our \model. The training pipeline described below is used to build this model, while online inference only requires the trained policy model to generate a label, confidence level, and reasoning trace. As illustrated in Fig.~\ref{fig:DeepScrub}(a), to address the lack of domain expertise in general LLMs, we perform domain knowledge injection via continued domain adaptation on a compliance-reviewed corpus composed of public references, governance materials, and de-identified case knowledge. To mitigate potential degradation of the model’s general capabilities caused by domain adaptation, we also incorporate general-domain corpora into the continued pretraining process. As shown in Tab.~\ref{knowledge}, a risk‑control to general data ratio of 1:15 achieves the best trade-off between domain adaptation and general capability preservation (C-Eval 73.97 vs. 73.87 baseline, essentially preserved; C-MMLU 75.80 vs. 76.93 baseline, a manageable 1.1-point drop). The resulting domain‑enhanced LLM serves as the base model for the subsequent SUggest–REflect (SURE) reinforcement learning stage.

The construction of the SURE training data is shown in Fig.~\ref{fig:DeepScrub}(b). 
Given heterogeneous multi-modal risk-assessment signals, we first apply standardized semantic templates to convert networks, sequences, and tabular records into unified textual descriptions, 
which are compatible with the input of LLMs. 
Data augmentation and filtering strategies are then employed to ensure label correctness and high data quality. 
The processed data are used to train the model in the SURE reinforcement learning stage, as depicted in Fig.~\ref{fig:DeepScrub}(c).

To address the limitation that existing RL methods lack structured reflective feedback for complex brushing detection, we introduce the SURE mechanism. By incorporating feedback from pre-trained risk-control expert and transaction‑expert models through a suggestion module, SURE guides the model to perform self-reflection and iterative refinement, thereby improving both decision quality and interpretability.

\subsection{Semantic Unification for Reasoning.}
\label{sub:COT}

\vpara{Multi-modal Semantic Unification}
As outlined in Sec.~\ref{sec:intro}, fake-order fraud detection requires processing multi-modal data, which presents a modality gap for text‑based LLMs. To bridge this gap, we define a standardized transformation template that converts heterogeneous data into LLM‑compatible text:

\vvpara{$\bullet$ Graph-to-Text. }
Entity association signals based on de-identified linkage evidence form a graph structure, which is valuable for identifying group risks and enhancing the performance of brushing detection. 
Thus, we convert the graph structure into a textual format to help LLMs better interpret these important relationships. 

\vvpara{$\bullet$ Sequence-to-Text. }
Sequential interaction patterns relevant to risk assessment represent their operational trajectory and risk tendencies, which are crucial for identifying bot-like behavior. 
We transform de-identified behavior-derived signals into textual representations for downstream reasoning. 
To enhance the information density of the sequence data, we further aggregate repeated actions in long sequences and summarize action patterns, ensuring high-quality data. 

\vvpara{$\bullet$ Table-to-Text. }
For structured records, aggregate indicators relevant to risk assessment are converted into textual summaries. We also apply a template to convert table data into text. Specifically, the feature names and their corresponding values are combined to create complete textual descriptions, while boolean and enumerated variables are transformed into their natural-language descriptions, rather than merely listing numeric values. 

\begin{figure}[t]
  \centering
  \includegraphics[width=\linewidth]{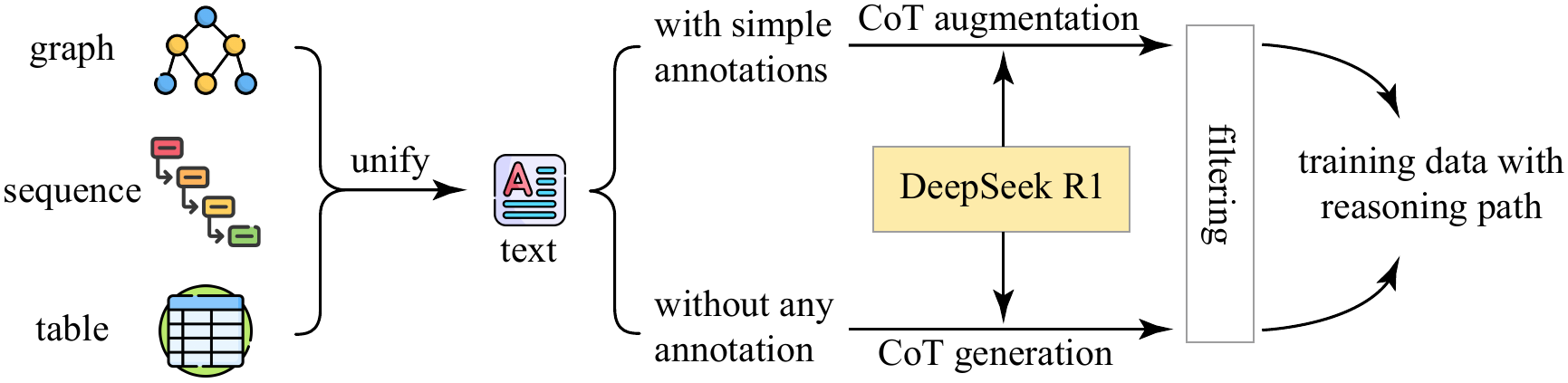}
  \caption{
  Construction of reasoning-augmented training data. Heterogeneous order evidence is first converted into text, then expanded with CoT augmentation or CoT generation, and finally filtered to keep label-consistent reasoning paths.
  }
  \label{fig:data_generation}
\end{figure}

After these transformations, heterogeneous multi-modal transaction data is unified into a text format that is more easily understood by the LLM, which will then undergo further data augmentation. Fig.~\ref{fig:data_generation} illustrates how we construct reasoning-augmented training data. Orders with brief manual analyses are expanded through chain-of-thought (CoT) enhancement, where DeepSeek-R1~\cite{guo2025deepseek} rewrites expert notes into structured reasoning paths. Orders without rationales are processed through CoT generation conditioned on their labels. We then filter generated rationales through manual verification to reduce label-inconsistent reasoning.

\vpara{CoT Data Amplification.}
Annotating fake-order transactions is highly dependent on domain experts, leading to high costs, limited scalability, and a lack of interpretable reasoning in existing labels. To address this issue, we have designed two data augmentation processes to construct training samples that achieve both broad coverage and high quality.

\vvpara{$\bullet$ CoT Enhancement.}
Expert annotations are often brief verdicts (e.g., ``suspected brushing'')  lacking step‑by‑step logic. We use DeepSeek‑R1 to reconstruct them into coherent reasoning chains that retain expert insight while adding explicit logical structure, enhancing both readability and support for training \model’s reasoning capability. The detailed prompt design is provided in App.~\ref{prompt:CoT Augmentation}.

\vvpara{$\bullet$ CoT Generation.}
For the large volume of rationale‑free samples, we employ a generation strategy conditioned on labels, which guides the model to produce reasonable reasoning paths aligned with ground truth annotations. The corresponding prompt templates and generation strategies are detailed in App.~\ref{prompt:CoT Generation}.

After completing the above data augmentation, all transaction data in textual form are equipped with labels accompanied by high-quality CoT rationales. To ensure consistency between the augmented or generated reasoning chains and the true labels, we further conduct manual verification to filter out noisy data.

\begin{figure*}[htp!]
  \centering
  \includegraphics[width=0.75\linewidth]{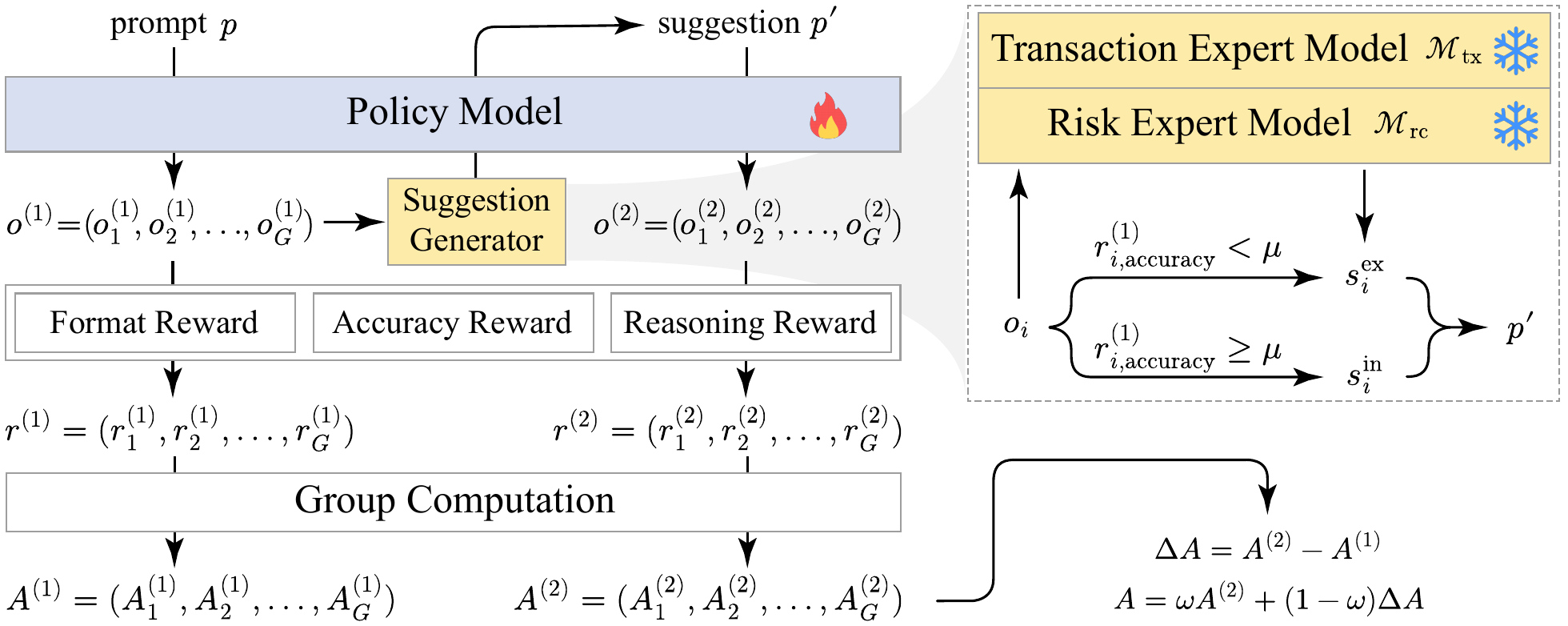}
  \vspace{-3mm}
  \caption{
  SURE training mechanism. The policy model first generates candidate responses, receives intrinsic or expert suggestions based on reward signals, and then refines its reasoning through a second-round response.
  }
  \label{fig:sure}
\end{figure*}

\subsection{Reward Modeling for Fake-order Fraud Detection}
\label{sec:reward}
Rewards guide model training by shaping its optimization direction. For reasoning‑based brushing detection, it is essential to evaluate both the final decision's correctness and the reasoning process's quality. However, existing work has largely overlooked this requirement. To address this gap, we design three reward functions for brushing detection: a format reward, an accuracy reward, and a reasoning reward. The detailed designs are described below.

\vvpara{$\bullet$ Format reward.}
Following the common practice in DeepSeek-R1, we introduce a format reward to encourage the model to present its analysis and conclusions in a structured manner. Specifically, the reasoning process is enclosed within the <reason> and </reason> tags, while the final decision is placed between the <answer> and </answer> tags. The resulting format reward is denoted as $r_{\text{format}}$.

\vvpara{$\bullet$ Accuracy reward.}
This reward measures both the correctness of the model’s prediction and its confidence in the predicted response. 
Concretely, when determining the fraud motivation of an order, the model is required to output an answer with an associated confidence level (e.g., ``<reason>...</reason><answer>Merchant inflates sales volume, confidence: high</answer>''). 
Let the model-generated reasoning process be $\hat{e} \in \mathbb{R}^{|\hat{e}|}$, the predicted answer be $\hat{y}$, the ground-truth reasoning process be $e$, the ground-truth answer be $y$, and the predicted confidence be $\hat{c}$, then we evaluate the model's judgment accuracy by combining answer correctness and confidence. 

First, we map the confidence level to a weight:
\begin{equation}
w(\hat{c}) = \begin{cases}
1,  & \ \text{if } \hat{c} = \text{‘high’} \\ 
2/3,& \ \text{if } \hat{c} = \text{‘medium’} \\ 
1/3,& \ \text{if } \hat{c} = \text{‘low’} \end{cases},
\end{equation}
and define the indicator function for answer correctness as

\begin{equation}
    \mathbb{I}(\hat{\bm{y}}, \bm{y}) = 
    \begin{cases} 
    1, & \text{if } \hat{\bm{y}} = \bm{y} \\ 
    0, & \text{if } \hat{\bm{y}} \neq \bm{y}
    \end{cases},
\end{equation}

Multiplying \(\mathbb{I}(\hat{y}, y)\) with confidence weight \(w(\hat{c})\) yields a basic accuracy signal, but this sparse reward may hinder training. To alleviate this issue, we further introduce a process-level score defined as the textual similarity between the model-generated reasoning \(\hat{e}\) and the ground-truth reasoning \(e\), allowing partial credit when the answer is wrong but the reasoning aligns with expert logic. The similarity is defined as:

\begin{equation}
    \ \text{sim}(\bm{\hat{e}},\ \bm{e})\ =\ \frac{f(\bm{\hat{e}})\ \cdot f(\bm{e})}{\|f(\bm{\hat{e}})\|\ \|f(\bm{e})\|}\ =\ \frac{\bm{u}\ \cdot \bm{v}}{\|\bm{u}\|\ \|\bm{v}\|}\ =\ \frac{\sum_{i=1}^d\ u_i\ v_i}{\sqrt{\sum_{i=1}^d\ u_i^2}\ \sqrt{\sum_{i=1}^d\ v_i^2}}\ ,
\end{equation}

where \(f\) denotes a Sentence-BERT encoder~\cite{reimers2019sentence}, \(u, v \in \mathbb{R}^d\) are the encoded sentence embeddings,  
\(\|\cdot\|\) denotes the \(L_2\) norm, and \(d\) is the embedding dimension.

Finally, we compute the accuracy reward by taking a weighted sum of the above components:

\begin{equation}
    r_{\text{accuracy}} = (1 - \alpha) \cdot \mathbb{I}(\hat{y}, y) \cdot w(\hat{c}) + \alpha \cdot \text{sim}(\hat{e}, e).
\end{equation}

Here, \(r_{\text{accuracy}}\) denotes the resulting accuracy reward, and \(\alpha\) is a hyperparameter that controls the contribution of the process similarity score. In our experiments, we set \(\alpha = 0.1\).

\vvpara{$\bullet$ Reasoning reward.}
The reasoning reward is designed to measure the extent to which the reasoning process contributes to deriving the correct answer, encouraging the model to produce not only correct predictions but also coherent and meaningful reasoning paths. The underlying intuition is that a model response consists of two components—the reasoning process \(\hat{e} \in \mathbb{R}^{|\hat{e}|}\) and the final answer \(\hat{y}\). If \(\hat{e}\) is informative and helpful, then the conditional probability of generating the correct answer tokens following \(\hat{e}\) should be relatively high.

Formally, the probability of generating the ground-truth answer \(y\) conditioned on the reasoning process \(\hat{e}\) is defined as

\begin{equation}
    A(\hat{e}, y) = \frac{1}{|y|} \sum_{n=1}^{|y|} P_\theta\bigl( y_n \mid \text{concat}(\hat{e}, y_{<n}) \bigr),
\end{equation}

where \(\hat{e}\) denotes the reasoning process, \(y\) is the ground-truth answer, and \(|y|\) is the length of \(y\). A larger \(A(\hat{e}, y)\) indicates that the model is more confident in producing \(y\) after observing \(\hat{e}\). However, this quantity only captures the local contribution of the reasoning process. To assess its global contribution, we further introduce perplexity as a measure of the model’s overall uncertainty over the generated text. Specifically, we concatenate \(\hat{e}\) and \(y\) into a complete sequence \(a\) and compute its perplexity as

\begin{equation}
    \mathrm{PPL}(a) = \exp\left( -\frac{1}{|a|} \sum_{n=1}^{|a|} \log P_{\theta}(a_n \mid a_{<n}) \right),
\end{equation}

where \(\bm{a} = \text{concat}(\hat{\bm{e}}, \bm{y})\) denotes the concatenated text.  
To combine these two metrics, we define a composite score \(\text{score}(\hat{\bm{e}}, \bm{y})\) that integrates both local and global perspectives on the usefulness of \(\hat{e}\):

\begin{equation}
    \text{score}(\hat{\bm{e}}, \bm{y}) = \frac{A(\hat{\bm{e}}, \bm{y})}{\text{PPL}(\bm{a})},
\end{equation}

To further reduce bias, we compute the difference between the score obtained with and without the reasoning process:

\begin{equation}
    \Delta\text{score} = \text{score}(\hat{\bm{e}}, \bm{y}) - \text{score}(\varnothing, \bm{y}),
\end{equation}

Moreover, to mitigate reward hacking from \(\Delta\text{score}\), we explicitly enforce a semantic connection between the reasoning process and the final answer. We extract a set of keywords \(\mathcal{K}(y)\) associated with each ground-truth answer and compute the proportion \(\gamma\) of these keywords that appear in the model-generated reasoning \(\hat{e}\):

\begin{equation}
    \gamma = \frac{1}{|\mathcal{K}(y)|} \sum_{k \in \mathcal{K}(y)} \mathbb{I}(k \in \hat{e}),
\end{equation}

where \(\mathbb{I}(\cdot)\) is the indicator function. The final reasoning reward is defined as 

\begin{equation}
r_{\text{reason}} = \Delta\text{score} \times \gamma.
\end{equation}

Here, \(r_{\text{reason}}\) denotes the reasoning reward, \(\gamma\) represents the proportion of answer-related keywords appearing in the reasoning process, and \(\mathcal{K}(y)\) denotes the keyword set associated with the ground-truth answer \(y\).

\subsection{Suggest-reflect Reinforcement Learning}
\label{sec:sure}
As noted in Sec.\ref{sec:intro}, existing RL methods lack the structured reflection needed for fraud reasoning. To overcome this, we propose a reinforcement learning framework with a SUggest–REflect mechanism (SURE). During training, SURE utilizes a pretrained transaction-expert model and a risk-control expert model to provide suggestions or trigger re-evaluation on the policy model’s intermediate responses. These suggestions drive the policy model to reflect on and reconstruct its outputs after incorporating the feedback. 

SURE aims to help the model better internalize the reasoning pathways underlying fraud prediction, and correct errors systematically. In doing so, SURE encourages the model to ``think like an expert'' in a systematic manner, thereby continuously improving both decision quality and interpretability in real-world deployments. As shown in Fig. \ref{fig:sure}, SURE uses an iterative training paradigm with dual-branch (intrinsic or extrinsic) suggestions. 

Specifically, transaction information is first organized into a prompt and fed into the policy model to generate an initial response. The suggestion module then produces intrinsic or extrinsic feedback via a dual‑branch evaluation. In the reflection stage, the model re‑examines its initial output together with the suggestions to produce a refined response. The advantages of the initial and final responses are then combined via a weighted average to form a mixed advantage signal, which is used to guide the policy update. The method details are as follows.

Formally, for each order, the transaction information \(\bm{t}\) is organized into a prompt \(p\) and fed into the policy model \(\pi_{\theta}\), which samples a group of first-round outputs \(\bm{o}^{(1)}=\{o_1^{(1)},o_2^{(1)},...,o_G^{(1)}\}\), where \(G\) denotes the number of sampled outputs. Each output \(o_i^{(1)}\) consists of a reasoning process \(\bm{\hat{e}}_i^{(1)}\) concatenated with a final answer \(\bm{\hat{y}}_i^{(1)}\), with \(i \in [1,G]\). Subsequently, the first-round responses \(\bm{o}^{(1)}\) are passed to the suggestion module.

\vvpara{$\bullet$ Suggest. }
Within the suggestion module, we first compute the reward scores for \(\bm{o}^{(1)}\) using the task-specific reward function introduced in Sec. \ref{sec:reward}, obtaining \(\bm{r}^{(1)}=\{{r}_\text{format}^{(1)},{{r}_\text{accuracy}^{(1)},{r}_\text{reason}^{(1)}}\}\). To prevent excessive deviation of the policy model’s own distribution caused by an overload of external suggestions (as also noted in \citep{yan2025learningreasonoffpolicyguidance}), we introduce a dual-branch decision mechanism. Specifically, for each first-round output \(\bm{o}_i^{(1)}\), a branch is selected based on its accuracy reward \(r_{i,\text{accuracy}}^{(1)}\). When the correctness reward is low (below a threshold \(\mu\)), it indicates that the model’s understanding is inadequate; in this case, we additionally invoke transaction‑expert and risk‑control‑expert models to provide extrinsic suggestions. Otherwise, if the correctness reward is sufficiently high, the model is considered to have produced a relatively reliable response, and only intrinsic suggestions are obtained to guide the model in self‑reflection.

\begin{equation}
    s_i = \begin{cases} 
    s_i^{\text{in}}, & \text{if the accuracy reward } r_{i,\text{accuracy}}^{(1)} \geq \mu, \\ 
    s_i^{\text{ex}}, & \text{otherwise}.
    \end{cases}
\end{equation}

Here, \(s_i^{\text{in}}\) denotes the intrinsic suggestion for output \(o_i^{(1)}\), which prompts the policy model to conduct self-reflection—checking the plausibility of its reasoning chain and exploring potentially better inference paths. Meanwhile, \(s_i^{\text{ex}}\) denotes the extrinsic suggestion for \(o_i^{(1)}\), which is provided by two external models: a transaction‑expert model and a risk-control-expert model. Specifically, the transaction-expert model compares the policy model’s generated reasoning \(\bm{\hat{e}}_i^{(1)}\) and answer \(\bm{\hat{y}}_i^{(1)}\) with the ground-truth reasoning \(\bm{e}\) and answer \(\bm{y}\), and suggests that the policy model focus on the misunderstood or overlooked key points in the transaction information. Similarly, the risk-control-expert model highlights anomalous features that could help improve the accuracy of fake-order fraud detection. Detailed prompt designs for these suggestions are provided in Appendix~\ref{appendix:prompts}.

\vvpara{$\bullet$ Reflect. }
In the subsequent reflection stage, for each first-round output \(o_i^{(1)}\), the suggestion content \(s_i\) together with the original output \(o_i^{(1)}\) are fed again into the policy model \(\pi_{\theta}\), prompting the model to examine its initial response and generate a refined final answer \(o_i^{(2)}\). Analogously, we compute the reward scores for the final responses \(\bm{o}^{(2)}\) using the same task-specific reward function introduced in Sec.\ref{sec:reward}, obtaining \(\bm{r}^{(2)}=\{{r}_\text{format}^{(2)},{{r}_\text{accuracy}^{(2)},{r}_\text{reason}^{(2)}}\}\).
Finally, the advantage of the two rounds of responses is combined via a weighted average to obtain the mixed advantage \(A^i\). Specifically, we introduce a hyperparameter \(\omega \in [0,1]\) to control the weighting between the second-round advantage \(A_i^{(2)}=\frac{r_i^{(2)}-\text{mean}(\bm{r}^{(2)})}{\text{std}(\bm{r}^{(2)})}\) and the advantage increment \(\Delta A_i=A_i^{(2)}-A_i^{(1)}\), as formulated below:

\begin{equation}
    A_i=\omega A_i^{({2})}+(1-\omega) \Delta A_i=\omega A_i^{(2)}+(1-\omega)(A_i^{(2)}-A_i^{(1)}).
\end{equation}

where \(A_i^{(1)}\) and \(A_i^{(2)}\) denote the advantage values of the first- and second-round responses \(o_i\), respectively; \(\Delta A_i\) measures the improvement in advantage after the suggestion–reflection stage; and \(A_i\) represents the blended advantage. This blended advantage is then used as the advantage term in the GRPO objective function to guide the update of the model parameters.

\section{Dataset Construction}

Our evaluation corpus is derived from a large-scale operational risk-control environment over a one-year observation window. As detailed in Sec. \ref{sub:COT}, each record is transformed into a de-identified textual representation for analysis. The construction of the dataset does not rely on a single validation method; instead, it employs a multi-layered filtering and cross-validation process to ensure the accuracy of sample labeling. The specific steps are as follows:

\vpara{Acquiring Fraudulent Samples.} Samples associated with fake-order fraud are directly extracted from daily risk-control enforcement outcomes, which provide the most reliable source of positive labels in our setting. These enforcement outcomes fall into three categories: cases labeled by experts, detections based on high-precision rules, and orders confirmed through law enforcement.

\vpara{Acquiring Benign Samples.} Selected from orders with prohibitively high fraud costs or logically low fraud probability, including: high‑value purchases (e.g., appliances, luxury goods), orders with extensive pre‑purchase consultation or reviews after consumption, and orders from users with long registration and rich purchase history.

\vpara{Data Cleaning and Cross-validation.} After obtaining the initially selected benign samples and fake-order fraud samples, we further process the data to exclude potential labeling errors. First, we remove samples with ambiguous labels. We then check feature consistency by filtering benign samples that exhibit patterns indicative of fake-order fraud and re-examining positive samples that closely resemble normal orders.

We evaluate \model on a real-world dataset collected from a major O2O e-commerce platform. The dataset contains approximately 150K de-identified orders and covers three categories: Merchant Inflating Product Sales, Influencer Inflating Their Level, and Normal Transaction. The raw class distribution is highly imbalanced. To better reflect deployment conditions and reduce temporal leakage, we split the data chronologically, using roughly the earlier 80\% for training and the later 20\% for testing. After undersampling the majority class in training, the class ratio is approximately 3 (Merchant) : 1 (Influencer) : 50 (Normal).

\begin{table*}[h]
\caption{Experimental results on our dataset for fake-order fraud detection.}
\vspace{-3mm}
\label{exp}
\centering
\begin{tabular}{lcccccccccccc}
\toprule
 \multicolumn{1}{c}{\multirow{2}{*}{\diagbox{Methods}{Metrics}}} & \multicolumn{3}{l}{\begin{tabular}[c]{@{}c@{}}Influencer Inflating\\ Their Level\end{tabular}} & \multicolumn{3}{l}{\begin{tabular}[c]{@{}c@{}}Merchant Inflating \\ Product Sales\end{tabular} } & \multicolumn{3}{c}{Normal Transaction}        & \multicolumn{3}{c}{Overall (macro)}           \\
\cmidrule(lr){2-4} \cmidrule(lr){5-7} \cmidrule(lr){8-10} \cmidrule(lr){11-13}
 & Prec.        & Rec.          & F1              & Prec.        & Rec.          & F1              & Prec.     & Rec.        & F1            & Prec.     & Rec.        & F1            \\
\midrule
Qwen3-32B~\cite{yang2025qwen3technicalreport}                   & 0.858          & 0.866          & 0.862          & 0.654          & 0.710          & 0.681          & 0.789          & 0.722          & 0.754          & 0.767          & 0.766          & 0.765          \\
Qwen3-8B~\cite{yang2025qwen3technicalreport} (SFT)                    & 0.905          & 0.923          & 0.914          & 0.759          & 0.719          & 0.739          & 0.814          & 0.835          & 0.824          & 0.826          & 0.826          & 0.826          \\
InternLM3-8B~\cite{cai2024internlm2} (SFT)               & 0.909          & 0.791          & 0.846          & 0.702          & 0.543          & 0.612          & 0.650          & 0.860          & 0.740          & 0.754          & 0.731          & 0.733          \\
Llama-3-8B~\cite{dubey2024llama} (SFT)                 & \textbf{0.918} & 0.915          & 0.916          & 0.752          & 0.727          & 0.739          & 0.813          & 0.826          & 0.819          & 0.827          & 0.823          & 0.825          \\
GLM-4-9B~\cite{glm2024chatglm} (SFT)                      & 0.906          & 0.928          & 0.917          & 0.769          & 0.687          & 0.725          & 0.801          & 0.854          & 0.826          & 0.825          & 0.823          & 0.823          \\
\midrule
\model w/o RL            & 0.913          & 0.915          & 0.914          & 0.772          & 0.719          & 0.744          & \underline{0.826}    & 0.834          & 0.830          & 0.837          & 0.823          & 0.830          \\
\model w/o SURE            & 0.909          & 0.932          & 0.920          & 0.780          & \underline{0.732}    & \underline{0.755}    & 0.822          & 0.850          & 0.836          & 0.837          & 0.838          & 0.837          \\
\model w/o $r_\text{accuracy}$  & 0.883          & 0.921          & 0.902          & 0.710          & 0.698          & 0.704          & 0.800          & 0.750          & 0.774          & 0.798          & 0.790          & 0.793          \\
\model w/o $r_\text{reasoning}$ & \underline{0.914}    & \underline{0.938}    & \underline{0.926}    & \underline{0.782}    & 0.731          & \underline{0.755}    & \textbf{0.832} & \textbf{0.871} & \textbf{0.851} & \underline{0.843}    & \underline{0.846}    & \underline{0.844}    \\
\model                     & \underline{0.914}    & \textbf{0.959} & \textbf{0.936} & \textbf{0.800} & \textbf{0.751} & \textbf{0.774} & \textbf{0.832} & \underline{0.869}    & \underline{0.850}    & \textbf{0.849} & \textbf{0.860} & \textbf{0.853} \\
\bottomrule
\end{tabular}
\end{table*}

\section{Experiment}
This section evaluates the effectiveness of our \model by addressing the following research questions:

\vvpara{$\bullet$Q1}: Does \model improve fake-order fraud review accuracy over baseline models? (Sec.~\ref{exp_res})

\vvpara{$\bullet$Q2}: Which components are responsible for the performance gains?
(Sec.~\ref{ablation})

\vvpara{$\bullet$Q3}: How does SURE improve reasoning quality through expert suggestions and self-verification? (Sec.~\ref{case})


\subsection{Experimental Setup}

\vpara{Training.} Based on the domain‑enhanced model (Sec.~\ref{sub:overview}), we perform supervised fine‑tuning (SFT) on 20\% of the CoT‑augmented training set (Sec.~\ref{sub:COT}) and reinforcement learning on the remaining 80\%. Optimization employs a learning rate of \(1\times10^{-5}\), a warmup ratio of 0.05, and a group size of 6. All experiments are run on 4 computing nodes with 32 high-performance GPUs, using a global batch size of 16 and a total training time of approximately 108 hours.
Key training hyperparameters are set as follows: reflection threshold $\mu = 0.5$, mixing weight between learning rounds $\omega = 0.7$.
The training pipeline is implemented using the ms‑swift~\cite{zhao2024swiftascalablelightweightinfrastructure} framework.

\vpara{Downstream Tasks.} The task is to classify orders into normal transaction, merchant inflating product sales, or influencer inflating their level, providing an interpretable rationale and reasoning path. To prevent feature leakage, all orders in the test set are generated after those in the training set.
For the fake-order fraud detection task, we evaluate using precision, recall, and F1, balancing the imbalanced classes via random sampling for per‑label analysis and using macro‑averaging for overall performance. The same protocol is applied to all baselines and our method for fair comparison.

\vpara{Baselines.} In this study, to evaluate the effectiveness of the proposed approach and address the practical need for resource constrained deployment, we design a hierarchical baseline system. First, we compare a smaller, task‑optimized model (based on Qwen3‑8B~\citep{yang2025qwen3technicalreport}) that has undergone domain enhancement and reinforcement learning against a much larger, general‑purpose model (Qwen3‑32B~\cite{yang2025qwen3technicalreport}) that has not been fine‑tuned for the task, testing whether targeted optimization can surpass scale. Second, at a similar parameter scale, we select widely used open‑source models, including InternLM3-8B-Instruct\footnote{InternLM3-8B-Instruct is an updated release from the same team as InternLM2~\cite{cai2024internlm2}}, Llama3‑8B~\citep{dubey2024llama}, and GLM4‑9B~\citep{glm2024chatglm} and fine‑tune them on the same training data as our method, thereby comparing different architectures under a fair training setup. This baseline design directly examines the trade‑off between small, tuned models and large, generic ones, while ensuring the gains are attributable to our contributions.

\subsection{Experimental Results}
\label{exp_res}
Tab.~\ref{exp} summarizes the overall performance of \model and major baseline methods on the fake-order fraud detection task. \model achieves the best results across all three classification subtasks, with a macro‑averaged F1‑score of 0.853, substantially outperforming all baselines. In detail, for influencer's level inflation, a relatively well‑defined pattern, \model attains an F1 of 0.936, demonstrating its stability on clear fraud signals. The most challenging merchant sales inflation subtask sees \model reaching an F1 of 0.774, exceeding the best fine-tuned baseline (0.739) by 3.5 points, which confirms its capability to detect complex, concealed brushing behaviors. Meanwhile, on normal transaction recognition, \model also achieves the highest F1 among all baselines (0.850), indicating that improved fraud detection does not come at the cost of misclassifying legitimate orders. Notably, our optimized 8B-level model outperforms the larger-scale Qwen3-32B model, which lacks domain adaptation, showing that task‑specific optimization is more effective than simply scaling model size in professional risk‑control scenarios.

\begin{table*}[h]
\caption{Correcting misclassification via experts' suggestions.}
\vspace{-3mm}
\label{case1}
\small
\begin{tabular}{lll}
\toprule
\multicolumn{3}{p{16.5cm}}{\textbf{Question}: ... You need to reason step by step based on the order characteristics to optimize the transaction motive derived from the manual analysis ... Basic transaction information of the order ...} \\
\midrule
\multicolumn{3}{p{16.5cm}}{\textbf{First-round Response}: <reason> ... the focal entity shows strong association signals with multiple roles in the transaction chain ... The behavioral pattern exhibits an anomalous cycle of "concentrated ordering followed by no verification or consumption."...It is possible that the \textcolor{red}{user is artificially inflating the payment-based Gross Merchandise Volume (GMV) metric to meet assessment targets (where "GMV" carries high weight in the influencer-level evaluation system)}...</reason><answer>\textcolor{red}{Influencer inflating their level}.</answer>} \\
\midrule
\multicolumn{3}{p{16.5cm}}{\textbf{Transaction-expert's Suggestion}:...The direct \textcolor{blue}{association between the user's identity and the merchant is not identified}... It is necessary to re-examine the differences between the influencer's level assessment metrics and the merchant sales evaluation criteria...
\textbf{Risk-control-expert's Suggestion}:...Abnormal surge in transaction amount... The timing distribution does not align with the user's typical behavior patterns...} \\
\midrule
\multicolumn{3}{p{16.5cm}}{\textbf{Second-round Response}: <reason>...\textcolor{green!80!black}{The user acts as an entity associated with the merchant}... The merchant fabricates sales volume through linked accounts... However, the user's behavior aligns more closely with "merchant inflating product sales," indicating a merchant-driven operational logic...</reason><answer>\textcolor{green!80!black}{Merchant inflating product sales}.</answer>} \\  
\bottomrule
\end{tabular}
\end{table*}

\begin{table*}[h]
\caption{Refining correct reasoning through self‑verification. The Question part is the same as that of Tab.~\ref{case1}.}
\vspace{-3mm}
\label{case2}
\small
\begin{tabular}{lll}
\toprule
\multicolumn{3}{p{16.5cm}}{\textbf{First-round Response}: <reason> ...The entity's profile is strongly consistent with the associated merchant type (based on contextual alignment signals). Their transactional behavior was characterized by high frequency, uniformity, and coordinated group activity which exhibited clear signs of sales inflation. The objective is to fabricate "product popularity" data to artificially boost merchant sales performance... </reason><answer>\textcolor{green!80!black}{Merchant inflating product sales}.</answer>} \\
\midrule
\multicolumn{3}{p{16.5cm}}{\textbf{Self-verification}:...Your conclusion is correct. Please proceed with the following steps for self‑verification to confirm whether a \textcolor{blue}{more optimal reasoning path exists}...} \\
\midrule
\multicolumn{3}{p{16.5cm}}{\textbf{Second-round Response}: <reason>...Behavior-derived risk patterns, together with cross-context consistency checks \textcolor{green!80!black}{(association with workplace scenarios) and the merchant’s promotional context (Double 11 period, recent platform entry)} collectively indicate a brushing scenario where the merchant aims to inflate sales of "gas + car wash" vouchers...</reason><answer>\textcolor{green!80!black}{Merchant inflating product sales}.</answer>} \\  
\bottomrule
\end{tabular}
\end{table*}

\subsection{Ablation Study}
\label{ablation}
To validate the contribution of each proposed component, we conduct four structured ablation experiments: (1) \model w/o RL: removing the entire reinforcement learning phase (SFT‑only baseline); (2) \model w/o SURE: removing the entire SURE mechanism; (3) \model w/o $r_\text{accuracy}$: ablating the accuracy reward while keeping the format and reasoning rewards, and (4) \model w/o $r_\text{reasoning}$: excluding the reasoning reward while retaining the accuracy and format rewards. Compared to the SFT‑only baseline, \model improves macro‑F1 from 0.830 to 0.853, showing that RL enables policy refinement beyond static supervision. Keeping RL but disabling SURE still achieves a competitive macro‑F1 (0.837), yet performance on “merchant inflating product sales” drops (F1=0.755 vs. 0.774), underscoring the value of interactive reflection for learning nuanced fraud patterns. Removing the accuracy reward causes the steepest decline: macro‑F1 falls to 0.793, and “merchant sales inflation” F1 drops to 0.704, confirming that this reward is essential for distinguishing fraudulent from legitimate behaviors. Excluding the reasoning reward yields a marginal drop on the simpler ‘influencer-level inflation’ task (F1=0.926 vs. 0.936) and a larger drop on ‘merchant sales inflation’ (0.755 vs. 0.774), indicating that the reasoning reward helps maintain logical consistency in complex scenarios.

\subsection{Case Study}
\label{case}
Tables~\ref{case1} and~\ref{case2} illustrate the two roles of SURE. In Tab.~\ref{case1}, expert suggestions help the model correct a wrong first-round label by revisiting overlooked merchant-user relationships and transaction patterns. In Tab.~\ref{case2}, self-verification improves an already correct prediction by adding more specific business context and behavioral evidence. These examples show that SURE improves both error correction and reasoning quality, which supports review-stage decisions in fake-order fraud detection.

\section{Real-world Deployment}
\label{real_world}
To verify the practical value of \model, we deployed it in a production risk-review workflow for local-service orders. \model is not used as a full-traffic detector. It operates after lightweight rules recall suspicious orders and automates the first-stage review that was previously handled by standard human reviewers.

\vpara{Deployment Architecture.}
\model runs as an automated first-stage reviewer in the human-audit pipeline on a cluster of six high-performance GPUs. Its workflow follows four stages: (1) filtering suspicious orders (flagged by risk‑control rules) and preparing multimodal inputs; (2) converting graphs, sequences, and tables into text via Sec.~\ref{sub:COT}; (3) generating a fraud verdict with a reasoning trace; (4) escalating cases to experts based on a threshold.

\vpara{Pilot Evaluation.}
During a four-week online trial, \model processed approximately 10,000 orders initially flagged as potentially fraudulent by risk control rules. All orders were simultaneously reviewed by the standard first-stage reviewers used in the production workflow for comparison. These reviewers had routine training but were not senior fraud experts. Final evaluation labels were adjudicated by senior fraud experts using the same evidence available in the production review workflow. The results showed that \model achieved a precision of 91.8\% (vs. 75.2\% human) and a recall of 88.5\% (vs. 49.7\% human), representing improvements of 16.6 and 38.8 percentage points, respectively.

\vpara{Production Impact.}
The pilot confirms that \model provides stronger first-stage review performance than standard human review in both coverage and accuracy. It is currently in production, reducing the  first-stage manual review workload by 94\% while cutting the time to review a single order from over ten minutes to a few seconds, saving nearly one million RMB annually. Each verdict includes a traceable reasoning chain to support expert adjudication. These results show that \model supports production fraud review with higher accuracy, lower first-stage review workload, and traceable reasoning for expert adjudication.
\section{Conclusion}

We presented \model, an LLM-based framework for fake-order fraud detection in production O2O risk-review workflows. By combining semantic unification, domain adaptation, and SURE training, \model improves detection performance and produces reasoning traces for review. Future work will extend the framework to broader fraud categories, reduce inference cost, and strengthen human review for high-impact cases.

\bibliographystyle{ACM-Reference-Format}
\bibliography{reference}
\appendix

\clearpage
\appendix
\section{Appendix}
\subsection{Ethics and Privacy}
This study strictly adheres to ethical standards and robust data security protocols to ensure data integrity, transparency, and user privacy. The study is evaluated on de-identified, access-controlled records derived from a large-scale online service platform under internal governance controls. Only authorized researchers who have undergone a formal internal approval process can access this proprietary data, thus ensuring data confidentiality and platform security.

All research records were processed under internal governance procedures and restricted to de-identified, aggregate, and access-controlled representations. No direct personal identifiers were used in the research workflow. This data is used solely to develop and evaluate fraud detection methods within a privacy-preserving framework that complies with relevant data protection regulations and platform policies.

To further ensure the transparency and security of our methodology, we employ only large, open-source language models and publicly available reinforcement learning libraries throughout the training and evaluation process. We do not use any proprietary or closed-source model components, thereby eliminating the risks associated with reliance on commercial models and improving the reproducibility of our research.

\subsection{Semantic transformation example}
\vvpara{$\bullet$ Graph-to-Text. }
For example, the group graph structure can be transformed into: ``A de-identified entity group contains 17 linked accounts, with aggregated activity observed over the recent period...''

\vvpara{$\bullet$ Sequence-to-Text. }
The final representation of a de-identified interaction sequence might look like: 
``A representative sequence of platform events relevant to risk assessment was observed and summarized.''

\vvpara{$\bullet$ Table-to-Text. }
Aggregated activity indicators over a short observation window were summarized into structured textual evidence, together with contextual consistency signals.

\subsection{Evaluation Metrics}

\vvpara{$\bullet$ Precision}: Precision measures the proportion of true positive predictions among all instances predicted as positive. It is particularly important when the cost of false positives is high, as it reflects the model's ability to avoid erroneous positive classifications. A higher precision indicates more reliable positive predictions, while a lower value suggests a greater rate of false positives in the model's output.
    \[
    \text{Precision} = \frac{TP}{TP + FP},
    \]
where TP is the number of true positives and FP is the number of false positives.

\vvpara{$\bullet$ Recall}: Recall (also called sensitivity or true positive rate) quantifies the proportion of actual positive instances that the model correctly identifies. This metric is critical in scenarios where missing positive cases, such as in medical screening or fraud detection. A high recall indicates that the model captures the majority of relevant positive cases.
    \[
    \text{Recall} = \frac{TP}{TP + FN},
    \]
where TP is the number of true positives and FN is the number of false negatives.

\vvpara{$\bullet$ F1-score}: F1-score is the harmonic mean of precision and recall, which balances the trade-off between the two metrics by assigning equal weight to both. It is optimal for scenarios where false positives and false negatives are equally costly, providing a single comprehensive measure of model performance on positive classes.
    \[
    \text{F1-score} = 2 \cdot \frac{\text{Precision} \cdot \text{Recall}}{\text{Precision} + \text{Recall}} = \frac{2TP}{2TP + FP + FN},
    \]
where TP is the number of true positives, FP is the number of false positives, and FN is the number of false negatives.

\vvpara{$\bullet$ Macro-averaged Precision}: Macro-averaged Precision (Macro-precision) calculates the unweighted average of precision values across all classes, treating each class equally regardless of its sample size. This metric is suitable when all classes have equal importance, avoiding bias toward majority classes in imbalanced datasets.
    \[
    \text{Macro-precision} = \frac{1}{C} \sum_{i=1}^{C} \text{Precision}_i,
    \]
where $C$ is the total number of classes, and $\text{Precision}_i$ is the precision value of the $i$-th class.

\vvpara{$\bullet$ Macro-averaged Recall}: Macro-averaged Recall (Macro-recall) computes the unweighted average of recall values for all classes, giving equal consideration to each class's performance. It is critical for evaluating model ability to identify instances of all classes equally, especially in multi-class classification tasks with balanced class importance.
    \[
    \text{Macro-recall} = \frac{1}{C} \sum_{i=1}^{C} \text{Recall}_i,
    \]
where $C$ is the total number of classes, and $\text{Recall}_i$ is the recall value of the $i$-th class.

\vvpara{$\bullet$ Macro-averaged F1-score}: Macro-averaged F1-score (Macro-F1) is the unweighted average of F1-score values across all classes, balancing precision and recall equally for every class. It serves as a holistic multi-class performance metric that avoids favoring large classes, making it ideal for imbalanced multi-class classification with equal class priority.
    \[
    \text{Macro-F1} = \frac{1}{C} \sum_{i=1}^{C} F1_i,
    \]
where $C$ is the total number of classes, and $\text{F1-score}_i$ is the F1-score value of the $i$-th class.

\subsection{List of Prompt}
\label{appendix:prompts}

\begin{tcolorbox}[
    width=\textwidth,
    float*=t,
    title={Prompt of CoT Augmentation\label{prompt:CoT Augmentation}},  
    fonttitle=\bfseries\small,
    coltitle=black,
    colbacktitle=white,
    boxed title style={colframe=black!30, boxrule=0.5pt, arc=3pt},
    left=5pt,
    right=5pt,
    top=10pt,
    bottom=5pt,
    boxsep=3pt,        
]
\begin{lstlisting}
## Role and Task
You are an expert in analyzing transaction motives for an O2O platform. Your core task is to identify transaction motives by analyzing order characteristics. The user will provide order information, the corresponding transaction motive, and manual analysis records. You need to reason step by step based on the order characteristics to optimize the transaction motive derived from the manual analysis.

## Known descriptions of transaction motive scenarios
1.Merchant inflating product sales: A behavior of artificially inflating the product's visible sales volume at a minimal cost.
2.Influencer inflating their level: Increase the influencer's level by self-buying or hiring the black market to create fake orders.
3.Normal Transaction: A normal transaction refers to a purchase behavior where a consumer places an order, makes payment, and fulfills the obligation (such as in-store verification) or completes a compliant refund within the platform's established rules, all driven by genuine and spontaneous consumption needs.

## Analysis requirements
Follow the steps below based on the provided transaction motive and manual analysis. Reason out the possibilities step by step based on the characteristic data.Consider multi-dimensional characteristics such as time, space, and behavior patterns. Compare normal business fluctuations with abnormal patterns. Give a confidence assessment.

## Output format
    <reason>
        1. First point of analysis...
        2. Second point of analysis...
    </reason>
    
        The most likely motive scenario is: Scenario name
        Confidence level: [High/Medium/Low]
        Additional notes: [If any]
    

## Basic transaction information of the order
{{feature}}

## Manual analysis
{{manual_analysis}}

## Transaction motive
{{true_label}}
\end{lstlisting}
\end{tcolorbox}

\begin{tcolorbox}[
    width=\textwidth,
    float*=t,
    title={Prompt of CoT Generation\label{prompt:CoT Generation}},  
    fonttitle=\bfseries\small,
    coltitle=black,
    colbacktitle=white,
    boxed title style={colframe=black!30, boxrule=0.5pt, arc=3pt},
    left=5pt,
    right=5pt,
    top=10pt,
    bottom=5pt,
    boxsep=3pt        
]
\begin{lstlisting}
## Role and Task
You are an expert in analyzing transaction motives for an O2O platform. Your core task is to identify transaction motives by analyzing order characteristics.The user will provide order information and the corresponding transaction motive. You need to reason step by step based on the order characteristics to optimize the transaction  motive derived from the manual analysis.

## Known descriptions of transaction motive scenarios
1.Merchant inflating product sales: A behavior of artificially inflating the product's visible sales volume at a minimal cost.
2.Influencer inflating their level: Increase the influencer's level by self-buying or hiring the black market to create fake orders.
3.Normal Transaction: A normal transaction refers to a purchase behavior where a consumer places an order, makes payment, and fulfills the obligation (such as in-store verification) or completes a compliant refund within the platform's established rules, all driven by genuine and spontaneous consumption needs.

## Analysis requirements
Follow the steps below based on the provided transaction motive and manual analysis. Reason out the possibilities  step by step based on the characteristic data.Consider multi-dimensional characteristics such as time, space, and behavior patterns. Compare normal business fluctuations with abnormal patterns.Give a confidence assessment.

## Output format
    <reason>
        1. First point of analysis... 
        2. Second point of analysis...
    </reason>
    
        The most likely motive scenario is: Scenario name
        Confidence level: [High/Medium/Low]
        Additional notes: [If any]

## Basic transaction information of the order
{{feature}}

## Transaction motive
{{true_label}}
\end{lstlisting}
\end{tcolorbox}

\vspace{5mm}

\begin{tcolorbox}[
    width=\textwidth,
    float*=t,
    title={Prompt of Training\label{prompt:Training}},  
    fonttitle=\bfseries\small,
    coltitle=black,
    colbacktitle=white,
    boxed title style={colframe=black!30, boxrule=0.5pt, arc=3pt},
    left=5pt,
    right=5pt,
    top=10pt,
    bottom=5pt,
    boxsep=3pt        
]
\begin{lstlisting}
## Role and Task
You are an expert in analyzing transaction motives for an O2O platform. Your core task is to identify transaction motives by analyzing order characteristics. You need to reason step by step based on the order characteristics and finally output a precise judgment result of the transaction motive. 
## Known descriptions of transaction motive scenarios
1.Merchant inflating product sales: A behavior of artificially inflating the product's visible sales volume at a minimal cost. 
2.Influencer inflating their level: Increase the influencer's level by self-buying or hiring the black market to create fake orders.
3.Normal Transaction: A normal transaction refers to a purchase behavior where a consumer places an order, makes payment, and fulfills the obligation (such as in-store verification) or completes a compliant refund within the platform's established rules, all driven by genuine and spontaneous consumption needs.
## Analysis requirements
Reason out the possibilities step by step based on the characteristic data.Consider multi-dimensional characteristics such as time, space, and behavior patterns. Compare normal business fluctuations with abnormal patterns.Give a confidence assessment.
## Output format
    <reason>
        1. First point of analysis... 
        2. Second point of analysis...
    </reason>
    <answer>
        The most likely motive scenario is: Scenario name
        Confidence level: [High/Medium/Low]
        Additional notes: [If any]
    </answer>
## Basic transaction information of the order
{{feature}}
\end{lstlisting}
\end{tcolorbox}

\vspace{8mm}
\begin{tcolorbox}[
    width=\textwidth,
    float*=t,
    title={Prompt of Self-verification\label{prompt:self-ver}},  
    fonttitle=\bfseries\small,
    coltitle=black,
    colbacktitle=white,
    boxed title style={colframe=black!30, boxrule=0.5pt, arc=3pt},
    left=5pt,
    right=5pt,
    top=10pt,
    bottom=5pt,
    boxsep=3pt        
]
\begin{lstlisting}
## Task Description
Your conclusion is correct. However, to ensure the rigour of the reasoning, please self - verify according to the following steps:
    1.Sort out the key information in the question.
    2.Gradually check your logical derivation process and its rationality.
    3.Confirm whether there is a better solution path.
    4.Finally, re-affirm the conclusion and answer the order motive again according to the following format.
## Output format
    <reason>
        1. First point of analysis... 
        2. Second point of analysis...
    </reason>
    <answer>
        The most likely motive scenario is: Scenario name
        Confidence level: [High/Medium/Low]
        Additional notes: [If any]
    </answer>
\end{lstlisting}
\end{tcolorbox}

\begin{tcolorbox}[
    width=\textwidth,
    float*=t,
    title={Prompt of Reflection\label{prompt:reflection}},  
    fonttitle=\bfseries\small,
    coltitle=black,
    colbacktitle=white,
    boxed title style={colframe=black!30, boxrule=0.5pt, arc=3pt},
    left=5pt,
    right=5pt,
    top=10pt,
    bottom=5pt,
    boxsep=3pt        
]
\begin{lstlisting}
## Task Description
Please gradually check your current reasoning process and conclusions to identify possible logical loopholes, data biases, or insufficient business understanding.
1.For each step of reasoning, clarify:
    Whether the basis is sufficient (data, rules, business common sense)
    Whether the logic is rigorous (are there leaps or unvalidated assumptions?)
    Whether the conclusion covers key scenarios (are there exceptions not considered?)
2.The following are improvement suggestions provided by trade experts: {{domain_suggest}}; please analyze these suggestions item by item and clarify:
    Which suggestions can directly correct your reasoning loopholes? How to correct them?
    Which suggestions require additional data or verification? Are there priorities?
    Avoid directly copying the suggestions; instead, evaluate their applicability in combination with your original reasoning.
3.The following is the analysis of abnormal behaviors by risk experts: {{riskcontrol_suggestion}}; please compare your answer with the risk control expert's answer:
    Behavior coverage: Are there abnormal behavior patterns you did not notice?
    Risk weight: Did you underestimate the risk of certain behaviors?
    Difference explanation: If the conclusions are inconsistent, is it due to data differences, different business assumptions, or omissions in analysis dimensions?
4.Finally, re-answer the order motive based on the questions according to the following format:
    <reason>
        1. First point of analysis... 
        2. Second point of analysis...
    </reason>
    <answer>
        The most likely motive scenario is: Scenario name
        Confidence level: [High/Medium/Low]
        Additional notes: [If any]
    </answer>
\end{lstlisting}
\end{tcolorbox}

\begin{tcolorbox}[
    width=\textwidth,
    float*=t,
    title={Prompt of Suggestions from Transaction-expert\label{prompt:tr-ex}},  
    fonttitle=\bfseries\small,
    coltitle=black,
    colbacktitle=white,
    boxed title style={colframe=black!30, boxrule=0.5pt, arc=3pt},
    left=5pt,
    right=5pt,
    top=10pt,
    bottom=5pt,
    boxsep=3pt        
]
\begin{lstlisting}
## Role and Task
As an expert in O2O platform business, please conduct a strict comparative analysis of the business knowledge differences between the following student's answer and the reference answer. 
Note: Only point out the deficiencies or errors in professional knowledge in the student's answer and provide suggestions for improvement directions, but do not directly disclose the specific content of the reference answer.
## Input format
Student's answer: {{answer}}
Reference answer: {{true_answer}}
## Output requirements (strictly follow this format):
1.Knowledge deficiencies: List the specific business concepts, terms, or logics that are missing/incorrect in the student's answer (no more than 3 items).
2.Improvement suggestions: Provide suggestions for each deficiency
## Example output:
Knowledge deficiencies: Lack of analysis of key business indicators.
Improvement suggestions: It is recommended to systematically learn the platform's core indicator system.
\end{lstlisting}
\end{tcolorbox}

\begin{tcolorbox}[
    width=\textwidth,
    float*=t,
    title={Prompt of Suggestions from Risk-control-expert\label{prompt:rc-ex}},  
    fonttitle=\bfseries\small,
    coltitle=black,
    colbacktitle=white,
    boxed title style={colframe=black!30, boxrule=0.5pt, arc=3pt},
    left=5pt,
    right=5pt,
    top=10pt,
    bottom=5pt,
    boxsep=3pt        
]
\begin{lstlisting}
## Role and Task
You are a professional expert in analyzing brushing behavior. Please conduct a systematic analysis based on the provided features. Before giving the final judgment, gradually carry out the thinking process according to the following requirements to determine whether the given order has transaction abnormalities:
1.Feature analysis: Analyze the suspicious indicators of the input features item by item.
2.Correlation verification: Check the logical relevance between features.
3.Pattern comparison: Compare with common brushing behavior patterns.
4.Abnormality assessment: Identify features that do not conform to normal transaction rules.
## Output format requirements:
    <reason>
        1. State the analysis logic step by step. 
        2. Only speak based on the provided feature data.
        3. Mark key judgment basis.
    </reason>
    <answer>
        There is an abnormal transaction./There is not an abnormal transaction.
    </answer>
\end{lstlisting}
\end{tcolorbox}

\end{document}